# Spin polarization studies in half-metallic $Co_2TiX$ (X = Ge and Sn) Heusler alloys

Lakhan Bainsla* and K. G. Suresh[#]

*Department of Physics, Indian Institute of Technology Bombay, Mumbai 400076, India*

**Abstract**

In this paper, we investigated the $Co_2TiX$ (X = Ge, Sn) Heusler alloys by structural, magnetic and spin polarization measurements to probe the half-metallic nature. Alloys are synthesized using the arc melting technique, and found to exist in $L2_1$ crystal structure with Fm-3m space group. Curie temperature ($T_C$) is found to be 384 and 371 K for $Co_2TiGe$ and $Co_2TiSn$ respectively. The saturation magnetization ($M_S$) value of 1.8 and 2.0 $\mu_B$/f.u. are obtained at 5 K for for $Co_2TiGe$ and $Co_2TiSn$ respectively, compared to 2.0 $\mu_B$/f.u. calculated by Slater-Pauling rule. To obtain the spin polarization value, differential conductance curves are recorded at the ferromagnetic/superconducting point contact at 4.2 K. The current spin polarization ($P$) value of 0.63 ± 0.03 and 0.64 ± 0.02 are deduced for $Co_2TiGe$ and $Co_2TiSn$ respectively. Considering the high current spin polarization and $T_C$, these materials appear to be promising for spintronic devices.

------------------------------------------------------------------

Corresponding address:

[*]lakhan@iitb.ac.in

[#]suresh@phy.iitb.ac.in

## 1. Introduction

Ferromagnets with electrons of only one spin direction at the Fermi level give rise to highly spin polarized currents and hence useful in spintronic devices. Half-metallic ferromagnetic materials are considered to be most interesting because of high spin polarization present in them by virtue of 100% spin polarized band structure [1-2]. Among the half-metallic materials, Heusler alloys have a special place due to their higher Curie temperatures and tunable electronic structure compared to other materials [2,3]. In the large family of Heusler alloys, Co based alloys show relatively high Curie temperature and spin polarization [4-8]. Half-metallic ferromagnets (FM) have applications in spintronic devices as spin polarized current sources for magnetic tunnel junctions [9], current-perpendicular-to-plane giantmagnetoresistive (CPP-GMR) devices [10], spin injectors to semiconductors [6], and lateral spin valves [11].

In general, there are two types of Heusler alloys, namely full Heusler ($X_2YZ$) which exist in $L2_1$ crystal structure and half Heusler (XYZ) with $C1_b$ crystal structure. Here X and Y are the transition elements and Y is a non magnetic element. Recently the half-metallic quaternary Heusler alloys with equiatomic composition (XX'YZ) were also predicted and studied experimentally [5,6,12,13]. In this class of materials, each lattice site is occupied by a different element and hence they show rather different symmetry with LiMgPdSn prototype [12]. Many of the half-metallic Heusler alloys were studied in view of their possible applications in spintronic devices. $Co_2MnSi$ is one of the extensively studied ternary Heusler alloys with current spin polarization (*P*) value of 0.54 as revealed by the point contact Andreev reflection (PCAR) technique [8]. Recently another alloy, namely $Co_2FeSi$ attracted a lot of attention due to its high Curie temperature and half metallic nature with $0.45 < P < 0.61$ [14]. However, a strong degradation in the transport properties was observed at room temperature because the spin polarization decreases drastically with temperature [15,16]. Therefore further exploration of half-metallic Heusler alloys is strongly desired, both from the experimental and theoretical perspectives.

Recently, a high value of current spin polarization has been reported for $Co_2$ based Heusler alloys [4,7,8], which motivate us to study $Co_2TiX$ (X = Ge, Sn) series. It is to be mentioned here that $Co_2TiX$ (X = Si, Ge and Sn) alloys were studied earlier experimentally and theoretically in view of the half-metallic and thermoelectric properties [17], but there are no reports on the current spin polarization. These alloys were found to be half-metallic from ab-initio calculations [17,18]. Electrical resistivity with a cusp-like anomaly at $T_C$ was observed for these alloys. A high value of magnetoresistance [$MR = \{(R(H) - R(0))/R(0)\}*100$] was observed for Co$_2$TiSn (55 % at 306 K in 40 kOe), which make these materials very promising from the application point of view [19]. In the present work, the bulk alloys of ternary Co$_2$TiX (X = Ge and Sn) were synthesized and investigated by x-ray diffraction, magnetization and spin polarization measurements to probe of the half-metallic behavior. For the proceeding text in the manuscript, we refer Co$_2$TiX (X = Ge, Sn), Co$_2$TiGe and Co$_2$TiSn as CTX, CTG and CTS respectively.

## 2. Experimental Details

Polycrystalline bulk samples of CTX were prepared by arc melting the appropriate quantities of constituent elements in an inert atmosphere. To avoid any oxygen contamination, Ti ingot was melted before melting the sample. The ingots were flipped and melted five to six times to ensure the homogeneity. The melted ingots were sealed under high vacuum and a heat treatment of 21 days at 1073 K was done to further increase the homogeneity. The structural analysis of the alloys was done by performing the room temperature powder x-ray diffraction measurements in a Philips diffractometer using Cu-k$_\alpha$ radiation. Magnetization measurements were performed in a vibrating sample magnetometer (VSM) attached to a physical property measurement system (PPMS, Quantum Design). Thermomagnetic curves were recorded in both zero field cooled (ZFC) and field cooled warming (FCW) modes; in the ZFC mode sample was first cooled in the absence of any applied magnetic field and the data were recorded during warming. In FCW mode, the sample was cooled in presence of a field and data were recorded during warming. Spin polarization measurements were performed using the PCAR technique [20], in which the differential conductance curves were obtained at the point contact between a superconducting Nb tip and the sample for the low temperature measurements, both the sample and superconducting tip were immersed in a liquid He bath. Spin polarization of the conduction electrons for each junction was obtained by fitting the normalized differential conductance curves to the modified BTK model [21], using the spin polarization, superconducting band gap ($\Delta$) and the interfacial scattering parameter (*Z*) as the variables.

3. **Results and discussion**

Structural analysis of the samples was done by performing the Rietveld refinement of the powder diffraction data collected at room temperature, as shown in the Fig. 1. Both CTG and CTS are found to exist in $L2_1$ crystal structure with *Fm-3m* space group (# 225). The obtained lattice parameter values of 5.82 and 6.07 Å are in good agreement with the reported values [17]. It is known that electronic structure of the half-metallic Heusler alloys is very sensitive to the structural disorder [22]. In this context, the presence of the superlattice reflections (111) and (200) for both CTG and CTS indicating that the alloys exist in well ordered crystal structure, is an important aspect.

The temperature dependence of magnetization for both the alloys is shown in Fig. 2. The $\chi$ vs. *T* curves were obtained both under ZFC and FCW. Curie temperature ($T_C$) was calculated from the first derivative of magnetization (*M*) versus temperature (*T*) plot. $T_C$ values of 384 and 371 K are obtained for CTG and CTS respectively, while the reported values are 380 (CTG) and 355 K (CTS) [17]. We find considerable thermomagnetic irreversibility, which may be attributed to the pinning of the domain walls and such a behavior was also observed in $Co_2VGa$ [23]. We have also performed AC magnetization measurements to probe the possibility of any magnetically frustrated phase. However no frequency dependance was observed for $\chi'$ (real part of susceptibility) and $\chi''$ (imaginary part of susceptibility). The saturation magnetization ($M_S$) values of 1.8 and 2.0 $\mu_B$/f.u. are obtained at 5 K for CTG and CTS respectively as shown in Fig. 3. The value obtained for CTS is in close agreement with the Slater-Pauling value [24]. The $M_S$ value decreases with the temperature, and value of 1.4 and 1.6 $\mu_B$/f.u. were obtained at 300 K for CTG and CTS respectively.

The normalized differential conductance curves were recorded using PCAR technique at 4.2 K, as shown in Fig. 4. We assumed $\Delta = \Delta_1 = \Delta_2$ for the fitting due to absence of the proximity effect in the PCAR data. In the presence of proximity effect, $\Delta_1$ and $\Delta_2$ are the gap values for the Andreev reflection process ($\Delta_1$) and quasiparticle transport ($\Delta_2$). The shape of the conductance curves depends on the value of $Z$, curves become flat near $\Delta$ for low $Z$ values while a peak appears for high $Z$ values. The values of the best fitting (smallest $\chi^2$ value) parameters are given in the figures (a) and (b). The obtained values of the $\Delta$ are lower than that of the bulk superconducting Nb (1.5 meV), which is attributed to the multiple contacts which give rise to the suppression of the band gap [25]. The intrinsic value of the current spin polarization can be obtained by recording the differential conductance curves at $Z = 0$; however the lowest possible values are 0.23 and 0.12 for CTG and CTS respectively as shown in the Fig. 4(a) and 4(b). To deduced the intrinsic value of $P$, the $P$ vs. $Z$ plots were extrapolated down to $Z = 0$. The current spin polarization values thus estimated are $0.63 \pm 0.03$ and $0.64 \pm 0.02$ for CTG and CTS respectively, as shown in Fig.4(c) and 4(d). The deduced spin polarization values are higher than those reported for many half-metallic ternary and pseudo-ternary Heusler alloys [7,8, 14, 26-27].

It is to be noted that the spin polarization measured by the PCAR technique is the transport spin polarization and for a ballistic point contact, this can be expressed as the imbalance in the majority and minority spin currents as $P = [ N_\uparrow(E_F) v_{F\uparrow} - N_\downarrow(E_F) v_{F\downarrow}] / [N_\uparrow(E_F) v_{F\uparrow} + N_\downarrow(E_F) v_{F\downarrow}]$ [28], which is the most realistic and relevant parameter from application point of view. However, the actual spin polarization ($P_a$) of density of states can also be expressed as, $P_a = [N_\uparrow(E_F) - N_\downarrow(E_F)]/ [N_\uparrow(E_F) + N_\downarrow(E_F)]$. Here $N_{\uparrow(\downarrow)}(E_F)$ and $v_{F\uparrow(\downarrow)}$ are the DOS at the Fermi level and the Fermi velocities for spin up (down) electrons respectively. The values of $P$ and $P_a$ are same when the Fermi velocities of both the spin currents are equal [21, 28].

## 4. Conclusions

The ternary $Co_2TiX$ (X = Ge and Sn) alloys are synthesized by arc melting and investigated by x-ray diffraction, magnetization and spin polarization measurements to look for the half-metallic behavior. Both the alloys are found to crystallize in $L2_1$ crystal structure with Fm-3m space group (#225). $T_C$ values of 384 K and 371 K are obtained for CTG and CTS respectively. The saturation magnetization are found to be 1.8 and 2.0 $\mu_B$/f.u. at 5 K for CTG and CTS respectively. Out of these, the value obtained for CTS is in close agreement with value predicted by the Slater-Pauling rule. High values of current spin polarization of $0.63 \pm 0.03$ and $0.64 \pm 0.02$ are deduced for CTG and CTS respectively. Considering the high values of $T_C$ and the current spin polarization, these materials seem to be promising for spintronic devices.

**Acknowledgments**

One of the authors, L. Bainsla, would like to thank U.G.C., Government of India, for granting a senior research fellowship (SRF). Authors also thank Prof. K. Hono for providing the PCAR facility at the Magnetic material unit, NIMS, Tsukuba, Japan. They are also grateful to Dr. Y. K. Takahashi from NIMS for the valuable discussions during the PCAR data analysis.

**References**

[1] C. Chappert, A. Fert & F. N. V. Dau, Nature Mat. 6, (2007) 813.

[2] C. Felser, G. H. Fecher, and B. Balke, Angew. Chem. Int. Ed. 46, (2007) 668.

[3] K. Inomata, N. Ikeda, N. Tezuka, R. Goto, S. Sugimoto, M. Wojcik and E. Jedryka, Sci. Technol. Adv. Mater. 9, (2008) 014101.

[4] B.S.D.Ch.S. Varaprasad, A. Srinivasan, Y.K. Takahashi, M. Hayashi, A. Rajanikanth, K. Hono, Acta Materialia 60, (2012) 6257.

[5] L. Bainsla, K. G. Suresh, A. K. Nigam, M. Manivel Raja, B.S.D.Ch.S. Varaprasad, Y. K. Takahashi, and K. Hono, J. Appl. Phys. 116, (2014) 203902.

[6] L. Bainsla, A. I. Mallick, M. Manivel Raja, A. K. Nigam, B.S.D.Ch.S. Varaprasad, Y. K. Takahashi, Aftab Alam, K. G. Suresh and K. Hono, Phys. Rev. B 91, (2015) 104408.

[7] T. M. Nakatani, A. Rajanikanth, Z. Gercsi, Y. K. Takahashi, K. Inomata and K. Hono, J. Appl. Phys. 102, (2007) 033916.

[8] A. Rajanikanth, Y. K. Takahashi and K. Hono, J. Appl. Phys. 101, (2007) 023901.

[9] Y. Du, B. S. D. Ch. S. Varaprasad, Y. K. Takahashi, T. Furubayashi and K. Hono, Appl. Phys. Lett. 103, (2013) 202401.

[10] Y. Sakuraba, M. Ueda, Y. Miura, K. Sato, S. Bosu, K. Saito, M. Shirai, T. J. Konno, and K. Takanashi, Appl. Phys. Lett. 101, (2012) 252408.

[11] Ikhtiar, S. Kasai, A. Itoh, Y. K. Takahashi, T. Ohkubo, S. Mitani and K. Hono, J. Appl. Phys. 115, (2014) 173912.

[12] K. Özdogan, E Sasioglu and I. Galanakis, J. Appl. Phys. 113, 193903 (2013); X. Dai, G. Liu, G. H. Fecher, C. Felser, Y. Li, and H. Liu, J. Appl. Phys. 105, (2009) 07E901.

[13] V. Alijani, S. Ouardi, G. H. Fecher, J. Winterlik, S. S. Naghavi, X. Kozina, G. Stryganyuk, and C. Felser, Phy. Rev. B 84, (2011) 224416.

[14] L. Makinistian, M. M. Faiz, R. P. Panguluri, B. Balke, S. Wurmehl, C. Felser, E. A. Albanesi, A. G. Petukhov and B. Nadgorny, Phys. Rev. B 87, (2013) 220402(R); S. Yamada, K. Hamaya, T. Murakami, B. Varaprasad, Y. K. Takahashi, A. Rajanikanth, K. Hono, and M. Miyao, J. Appl. Phys. 109, (2011) 07B113.

[15] S. Li, Y. K. Takahashi, T. Furubayashi, and K. Hono, Appl. Phys. Lett. 103, (2013) 042405.

[16] Y. Sakuraba, M. Ueda, Y. Miura, K. Sato, S. Bosu, K. Saito, M. Shirai, T. J. Konno, and K. Takanashi, Appl. Phys. Lett. 101, (2012) 252408.

[17] J. Barth, G. H. Fecher, B. Balke, S. Ouardi, T. Graf, Claudia Felser, A. Shkabko, A. Weidenkaff, P. Klaer, H. J. Elmers, H. Yoshikawa, S. Ueda and K. Kobayashi, Phys. Rev. B 81, (2010) 064404.

[18] S. C. Lee, T. D. Lee, P. Blaha and K. Schwarz, J. Appl. Phys. 97, (2005) 10C307.

[19] J. Barth, G. H. Fecher, B. Balke, T. Graf, A. Shkabko, A. Weidenkaff, P. Klaer, M. Kallmayer, H. J. Elmers, H. Yoshikawa, S. Ueda, K. Kobayashi and C. Felser, *Phil. Trans. R. Soc. A* 369, (2011) 3588-3601.

[20] R. J. Soulen, J. M. Byers, M. S. Osofsky, B. Nadgorny, T. Ambrose, S. F. Cheng, P. R. Broussard, C. T. Tanaka, J. Nowak, J. S. Moodera, A. Barry, J. M. D. Coey, Science 282, (1998) 85.

[21] G. J. Strijkers, Y. Ji, F. Y. Yang, C. L. Chien, J. M. Byers, Phys. Rev. B 63, (2001) 104510.

[22] S. Picozzi, A. Continenza, and A. J. Freeman, Phys. Rev. B 69, (2004) 094423.

[23] T. Kanomata, Y. Chieda, K. Endo, H. Okada, M. Nagasako, K. Kobayashi, R. Kainuma, R. Y. Umetsu, H. Takahashi, Y. Furutani, H. Nishihara, K. Abe, Y. Miura, M. Shirai, Phys. Rev. B 82, (2010) 144415.

[24] I. Galanakis, P. H. Dederichs, N. Papanikolaou, Phys. Rev. B 66, (2002) 174429.

[25] S. K. Clowes, Y. Miyoshi, O. Johannson, B. J. Hickey, C. H. Marrows, M. Blamire, M. R. Branford, Y. V. Bugoslavsky, L. F. Cohen, J. Magn. Magn. Mater. 272, (2004) 1471.

[26] S.V. Karthik, A. Rajanikanth, Y.K. Takahashi, T. Ohkubo, K. Hono, Acta Materialia 55, (2007) 3867.

[27] A. Rajanikanth, D. Kande, Y. K. Takahashi and K. Hono, J. Appl. Phys. 101, (2007) 09J508.

[28] I. I. Mazin, Phys. Rev. Lett. 83, (1999) 1427.

**Figures:**

Figure 1

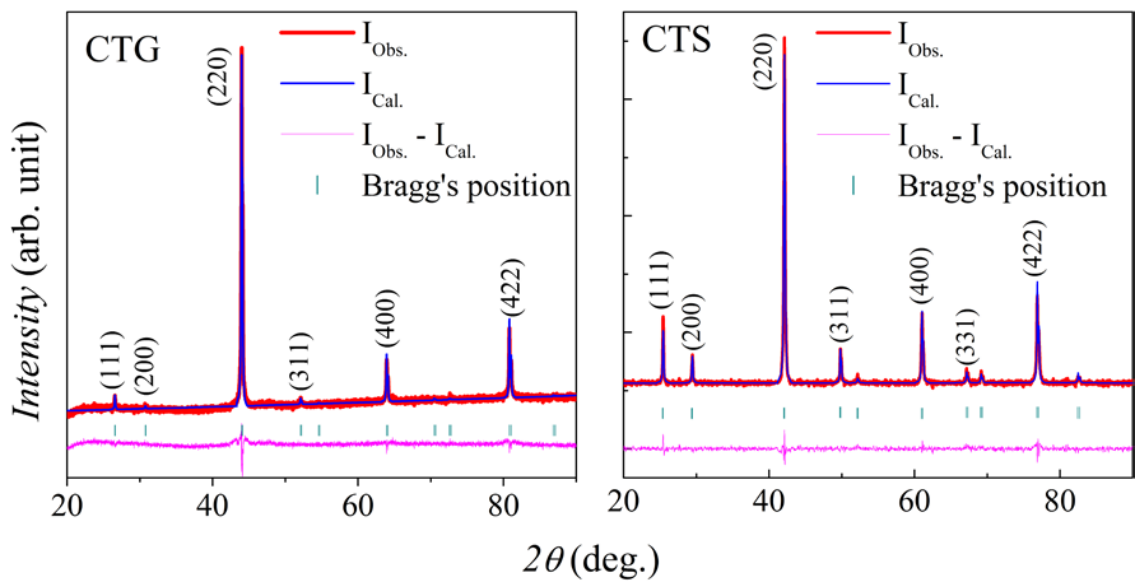

**Fig. 1.** Rietveld refinment of the room temperature powder x-ray diffraction pattern collected using Cu-K$_\alpha$ radiation.

Figure 2

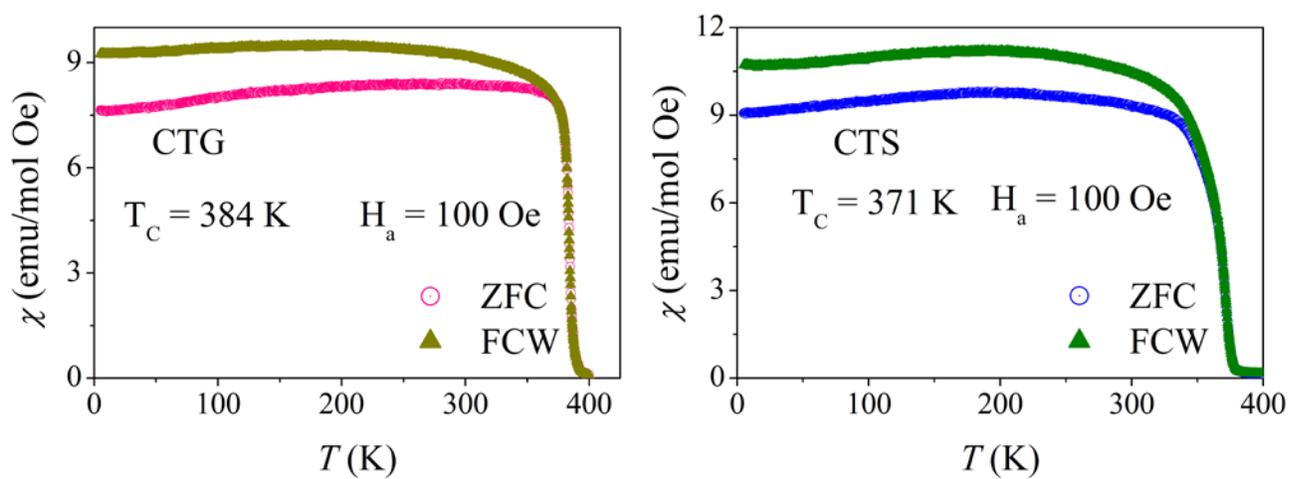

**Fig. 2.** Thermomagnetic curves obtained under an applied field of 100 Oe.

Figure 3

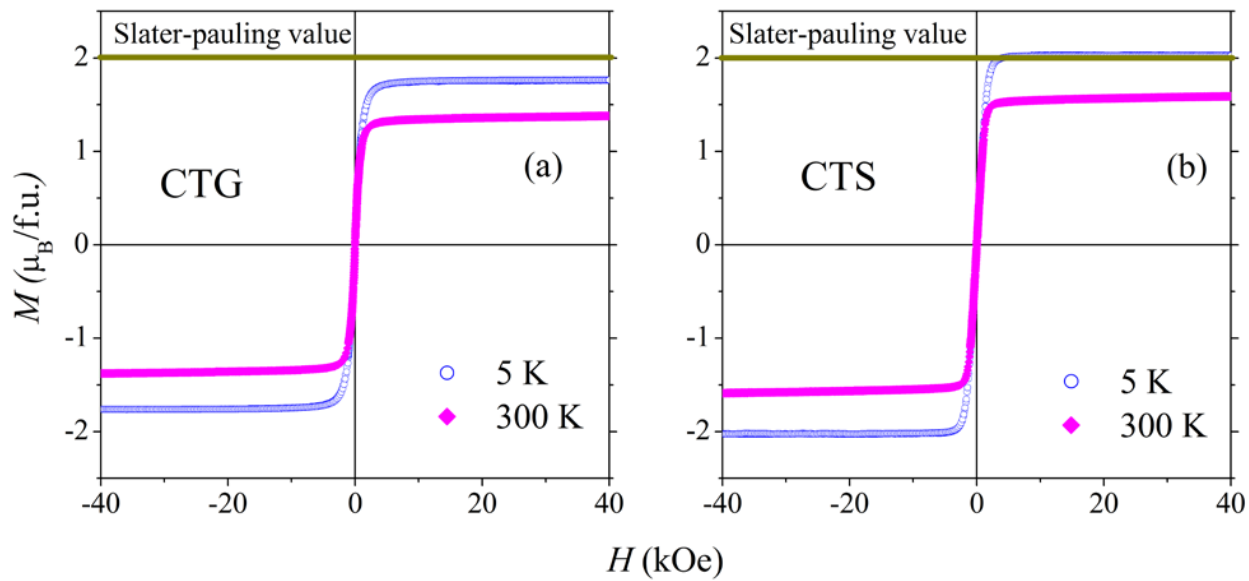

**Fig. 3.** Isothermal magnetion curves obtained at 5 K and 300 K in ± 40 kOe range. Open circles represent the data obtained at 5 K and the filled symbols refer to the data at 300 K. Solid lines refers to the value calculated by the Slater-Pauling rule.

Figure 4

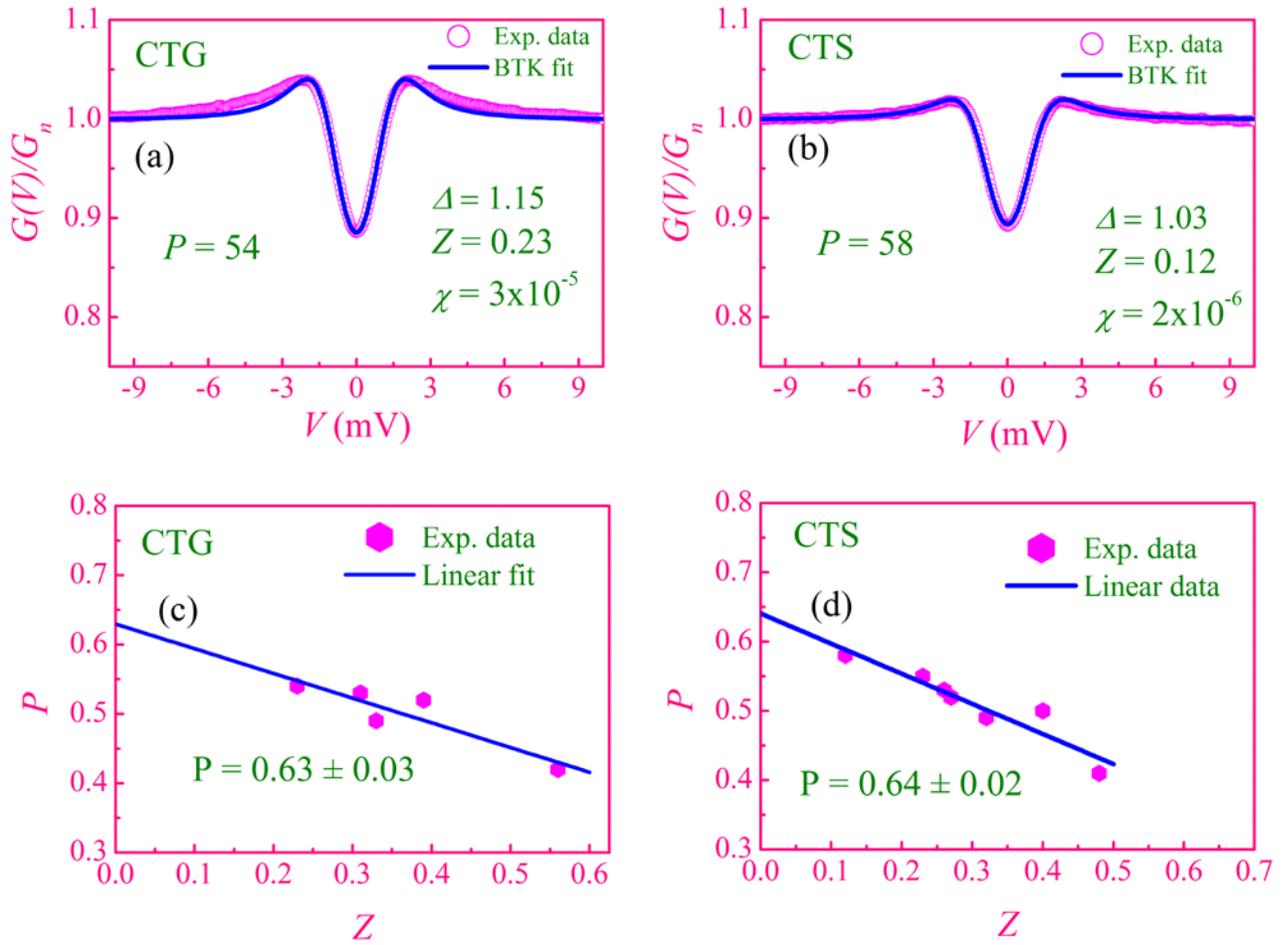

**Fig. 4.** (a) and (b) Normalized differential conductance curves with BTK fit for CTG and CTS respectively. The open circles represent the experimental data and the solid lines are the BTK fit to the data. (c) and (d): *P vs. Z* curves for CTG and CTS respectively. Here the filled symbols represent the experimental data and the solid lines are the linear fit to the data.